# Spin Hall effect in a spin-1 chiral semimetal

Ke Tang,[1,2] Yong-Chang Lau,[3,4] Kenji Nawa,[1,5] Zhenchao Wen,[1,*] Qingyi Xiang,[1] Hiroaki Sukegawa,[1] Takeshi Seki,[3,4] Yoshio Miura,[1] Koki Takanashi,[3,4,6] and Seiji Mitani[1,2]

[1]National Institute for Materials Science (NIMS), Tsukuba 305-0047, Japan
[2]Graduate School of Pure and Applied Sciences, University of Tsukuba, Tsukuba 305-8577, Japan
[3]Institute for Materials Research, Tohoku University, Sendai 980-8577, Japan
[4]Center for Spintronics Research Network, Tohoku University, Sendai 980-8577, Japan
[5]Graduate School of Engineering, Mie University, Tsu 514-8507, Japan
[6]Center for Science and Innovation in Spintronics, Core Research Cluster, Tohoku University, Sendai 980-8577, Japan

**Abstract:**

Spin-1 chiral semimetal is a new state of quantum matter hosting unconventional chiral fermions that extend beyond the common Dirac and Weyl fermions. *B*20-type CoSi is a prototypal material that accommodates such an exotic quasiparticle. To date, the spin transport properties in the spin-1 chiral semimetals, have not been explored yet. In this work, we fabricated *B*20-CoSi thin films on sapphire *c*-plane substrates by magnetron sputtering and studied the spin Hall effect (SHE) by combining experiments and first-principles calculations. The SHE of CoSi using CoSi/CoFeB/MgO heterostructures was investigated via spin Hall magnetoresistance and harmonic Hall measurements. First-principles calculations yield an intrinsic spin Hall conductivity (SHC) at the Fermi level that is consistent with the experiments and reveal its unique Fermi-energy dependence. Unlike the Dirac and Weyl fermion-mediated Hall conductivities that exhibit a peak-like structure centering around the topological node, SHC of *B*20-CoSi is odd and crosses zero at the node with two antisymmetric local extrema of opposite sign situated below and above in energy. Hybridization between Co *d*-Si *p* orbitals and spin-orbit coupling are essential for the SHC, despite the small (~1%) weight of Si *p*-orbital near the Fermi level. This work expands the horizon of topological spintronics and highlights the importance of Fermi-level tuning in order to fully exploit the topology of spin-1 chiral fermions for spin current generation.

*E-mail: wen.zhenchao@nims.go.jp

Spin current generated from the spin Hall effect (SHE) can exert spin-orbit torque (SOT) on an adjacent ferromagnetic (FM) nanolayer, which provides a promising way to manipulate the magnetization of the FM layer for high-performance spintronic devices [1,2], such as SOT-magnetoresistive random access memories (MRAMs). For applications involving SOT-based devices, it is important to find materials exhibiting large damping-like spin Hall efficiency ($\xi_{DL}$), i.e., the efficiency of spin-charge conversion of SHE. Heavy 5$d$ transition metals, such as Pt, Ta, and W, are commonly known for possessing large $\xi_{DL}$ due to their strong spin-orbit interactions and optimum $d$-orbital filling [3–6]. Topological insulators (TIs), such as $Bi_2Se_3$ [7–10], were reported to show even larger $\xi_{DL}$, which arises from the spin-momentum locking of their surface states. However, several complications hinder the integration of the TIs to the state-of-the-art electronics: First, the relatively low melting point of these Bi-based TIs is not compatible with the CMOS processing. Second, the bulk-insulating TIs are generally too resistive to be incorporated into devices, such as SOT-MRAMs. Third, in a multilayer heterostructure, the surface states may be altered when a TI is in direct contact with a ferromagnet.

The emergence of topological Weyl semimetals (WSMs) provides a promising alternative for efficient spin-current generation in SOT devices, due to the topological nature of the bulk band structures and the relatively low resistivity of WSMs compared to TIs. The WSMs hold three dimensional linearly dispersive band crossing points, where the degeneracy is lifted by breaking either the inversion symmetry or the time-reversal symmetry or both [11]. First-principles calculations have predicted a large intrinsic SHE peak near the Weyl points for the TaAs family of WSMs, where the SHE is interpreted as the interplay of the large spin Berry curvature and spin-orbit coupling (SOC) near the Weyl nodes [12]. Recently, large SHE was observed in Weyl semimetal prototypes $WTe_2$ [13,14] and $Co_2MnGa$ [15,16].

The cobalt monosilicide CoSi that crystallizes in the *B*20 structure is a newly discovered topological semimetal featuring unconventional chiral fermions with no counterparts in high-energy physics [17]. Left panel of Fig. 1(a) shows an illustration of the unit cell of cubic *B*20-structured CoSi. The spin-1 chiral fermion holds a threefold degeneracy at the band-crossing node of a Dirac-like band and a flat band, which carries a topological charge of ±2 [17,18], as illustrated in the right panel of Fig. 1(a). The presence of spin-1 chiral fermions (schematic shown in the right panel of Fig. 1(a)) at the band-crossing points near the Fermi level with long Fermi arcs in the momentum space of CoSi were confirmed by angle-resolved photoemission spectroscopy experiments [19–21]. Fabrication of high-quality CoSi thin films and understanding the spin transport in these topological semimetal thin films are essential steps towards realizing novel spintronic devices exploiting the unique properties of unconventional chiral fermions. To date, there have been no report on the spin-transport properties of nanometer-scale CoSi thin films and it is not clear yet how does the spin-1 chiral electronic structure participate in the spin transport, e.g. spin current generation via SHE.

In this work, we fabricated CoSi thin films by magnetron sputtering and investigated the spin-transport properties by experiments and first-principles calculations. Structural characterization indicates that the CoSi films are polycrystalline and crystallizes in the *B*20-type structure. Spin Hall magnetoresistance (SMR) and harmonic Hall measurements show that the damping-like spin Hall efficiency of CoSi is ~3%, which is larger than the values reported in pure Co (~1%) [22] and Si (~0.01%) [23]. The first-principles calculations show that the amplitude of the spin Berry curvature depends on the *k*-point path and changes sign above and below the topological node hosting spin-1 chiral fermion, in contrast to a WSM. The spin Berry curvature contributes to the spin Hall conductivity (SHC) in the CoSi thin films and originates from the hybridization of *d-p* orbitals between Co and Si.

CoSi thin films were deposited on sapphire (Al$_2$O$_3$) *c*-plane substrates from a sintered CoSi alloy target. The composition of the CoSi films was confirmed to be 50.7:49.3 using inductively coupled plasma optical emission spectrometer (ICP-OES). The surface structure of the CoSi films was monitored by *in-situ* reflection high energy electron diffraction (RHEED) in the sputtering chamber. Figure 1(b) is a typical RHEED pattern for a 47-nm-thick CoSi film deposited at 550°C. The RHEED pattern shows arc-shape diffractions, indicating a polycrystalline surface. The crystal structure was further characterized by x-ray diffraction (XRD) with Cu Kα radiation and a monochromator. Figure 1(c) shows the XRD spectrum of substrate/CoSi (47)/Mg$_2$Al-O$_x$ (2) (thickness in nanometers), where Mg$_2$Al-O$_x$ serves as a capping layer to protect the CoSi surface from oxidization. In addition to a diffraction peak from the sapphire substrate, the peaks corresponding to (200), (210), and (211) crystal planes of *B*20-type CoSi can be clearly identified, which indicates the polycrystalline nature of the CoSi film without observable secondary phases. From the intensities of the diffraction peaks, it is also known that the CoSi film has a (210)-preferential texture. The lattice constant of the CoSi is estimated to be ~0.4435 nm. Next, a heterostructure consisting of substrate/CoSi ($t_{CoSi}$: 0~11)/Co$_{20}$Fe$_{60}$B$_{20}$ (CoFeB) (1)/MgO (2)/Ta (1) was deposited in order to evaluate the spin-transport properties of the CoSi film. The surface morphology of the stack layers was measured by atomic force microscopy (AFM). As shown in Fig. 1(d), a relatively flat surface with an average roughness ($R_a$) ~ 0.39 nm and a peak-to-valley (*P-V*) ~ 3.92 nm for a 1×1 μm$^2$ scan was achieved. The heterostructure with such a low roughness is essential for spin-transport measurements, which will be further elaborated in the following sections.

The multilayer stacks with the core structure of CoSi/CoFeB/MgO were patterned into Hall bars, as shown in Fig. 2(a). SMR was measured to evaluate $\xi_{DL}$ of the CoSi thin films. In non-magnetic/ferromagnetic (NM/FM) layered structures, when a charge current is applied, a spin current can be generated by SHE in the NM and it flows towards the FM. At the interface of the NM/FM, a part of the spin current is reflected back and it is converted to a charge current within the NM by the inverse SHE, resulting in a change of the longitudinal resistance ($R_{xx}$) in the heterostructure. The difference in the $R_{xx}$ between parallel and perpendicular configurations of the FM layer magnetization (***M***) and the spin-current polarization ($\sigma \mathbin{/\mkern-6mu/} y$) is referred to as the SMR [24,25]. Neglecting the imaginary part of the spin mixing conductance, $\xi_{DL}$ of the NM layer can be obtained from the NM layer thickness dependence of the SMR.

Figure 2(a) illustrates the experimental setup for the SMR measurement and the microscope image of a Hall bar. A magnetic field of 20 kOe was applied to saturate the magnetization of the ferromagnetic CoFeB layer. By rotating the magnetic field in *zy* plane, the parallel and perpendicular configurations of ***M*** and ***σ*** were achieved. Figure 2(b) shows the $R_{xx}$ change with the orientation of ***M*** for two Hall bar devices with $t_{CoSi}$ = 4.4 nm and 9.0 nm. The $\theta = 0°$ and 90° correspond to the perpendicular and parallel states between ***M*** and ***σ***, respectively. The SMR is derived by the ratio of the $R_{xx}$ change, which is given by $(R_{xx,\,\theta=90°} - R_{xx,\,\theta=0°})/R_{xx,\,\theta=0°}$. Because the ***M*** is always perpendicular to the charge current, there is no contribution of anisotropic magnetoresistance (AMR) to the resistance change. Assuming a transparent interface between CoSi and CoFeB, the SMR as a function of $t_{CoSi}$ can be fitted by the following equation derived from a drift-diffusion model [25–27],

$$\text{SMR} = -\xi_{DL}^2 \frac{\lambda}{t_{CoSi}} \frac{\tanh(t_{CoSi}/2\lambda)}{1+\alpha} \left[1 - \frac{1}{\cosh(t_{CoSi}/\lambda)}\right]. \qquad (1)$$

The $\xi_{DL}$ and the spin diffusion length ($\lambda$) are used as fitting parameters. The $\alpha$ is given by $\rho_{CoSi} t_{CoFeB}/\rho_{CoFeB} t_{CoSi}$, where $\rho_{CoSi}$, $\rho_{CoFeB}$, and $t_{CoFeB}$ represent the resistivity of CoSi, the resistivity and the thickness of CoFeB, respectively. Here, $t_{CoFeB}$ is 1 nm and $\rho_{CoFeB}$ is 200 $\mu\Omega\cdot$cm. The value of $\rho_{CoSi}$ is derived by fitting the inverse of sheet resistance of the film as a function of $t_{CoSi}$, as shown in Fig. 2(c). The inverse of the sheet resistance is defined as $(1/R_{xx}) \times (L/W)$, where $L = 25$ µm is the length and $W = 10$ µm is the width of the Hall bar. Thus, the slope of the linear fitting curve represents the inverse of the $\rho_{CoSi}$. We fitted the data in an appropriate range and estimated the $\rho_{CoSi}$ is around 341 μΩ·cm. The value of $\rho_{CoSi}$ is relatively large compared to the reported resistivity of a bulk CoSi single crystal [28], which is attributed to the carrier scattering at the surfaces and the grain boundaries of the polycrystalline thin film. After extracting the SMR of bilayers with different thicknesses of CoSi, we fitted the SMR as a function of $t_{CoSi}$ by the Eq. (1). The experimental data are well fitted, as shown in Fig. 2(d). The $\xi_{DL}$ and λ are obtained to be 3.4 ± 0.2 % and 4.4 ± 0.6 nm, respectively. The value of $\xi_{DL}$ is relatively large even for a material consisting of relatively light elements. Compared to pure Co ($\xi_{DL}$ ~ 1%) [22] and Si ($\xi_{DL}$ ~ 0.01%) [23], the result indicates that the orbital hybridization between Co and Si orbitals may play a significant role on the SHE in the CoSi, which is supported by the first-principles calculations discussed later.

The efficiency of spin-current generation from the CoSi thin films was further characterized by harmonic Hall measurements. The SOTs induced by spin current acting on a FM layer can be considered, in the quasistatic regime, as effective magnetic fields, which consist of an damping-like ($\boldsymbol{H}_{DL} \parallel \boldsymbol{M} \times \boldsymbol{\sigma}$) term and a field-like ($\boldsymbol{H}_{FL} \parallel -\boldsymbol{\sigma}$) term [29–34]. The effective fields modulate the orientation of $\boldsymbol{M}$ of the FM layer, resulting in a second harmonic Hall resistance. The efficiencies of damping-like ($\xi_{DL}$) and field-like ($\xi_{FL}$) torques can be extracted from the azimuthal magnetic field angular dependence of the first ($R_\omega$) and the second ($R_{2\omega}$) harmonic Hall resistances. A schematic illustration of the harmonic Hall measurement is shown in Fig. 3(a). A sinusoidal charge current was applied along the $x$ axis, and an external magnetic field $\boldsymbol{H}_{ext}$ was applied and rotated in the $xy$ plane, making an angle $\varphi$ with the current. The $R_\omega$ and $R_{2\omega}$ were measured as a function of $\varphi$ at various $H_{ext}$. The dependences of the first and the second harmonic Hall resistances on the angle $\varphi$ can be expressed by the following equations [29],

$$R_\omega = R_{AHE} \cos\theta + R_{PHE} \sin 2\varphi \sin^2\theta \qquad (2)$$

$$R_{2\omega} = -\left(R_{AHE}\frac{H_{DL}}{H_{ext}+H_k} + R_{cost} + R_{ONE}\right)\cos\varphi + \left(2R_{PHE}\frac{H_{FL}+H_{Oe}}{H_{ext}}\right)\cos 2\varphi \cos\varphi$$

$$= A\cos\varphi + B\cos 2\varphi \cos\varphi, \tag{3}$$

where $R_{AHE}$, $R_{PHE}$, $H_k$, and $H_{Oe}$, are the anomalous Hall resistance, the planar Hall resistance, the out-of-plane magnetic anisotropy field of CoFeB, and the Oersted field. $R_{cost}$ and $R_{ONE}$ describe $R_{2\omega}$ contributions of thermoelectric origin that are constant in field (e.g. due to the anomalous Nernst effect of CoFeB) and linear in field (e.g. due to the ordinary Nernst effect (ONE) of CoSi), respectively. The $R_{2\omega}$ is decomposed into two components, i.e., $A\cos\varphi$ and $B\cos 2\varphi\cos\varphi$.

Figure 3(b) shows the dc Hall resistance ($R_{xy}$) of Hall devices with $t_{CoSi}$ = 4.5, 5.8, 7.2, and 8.5 nm. The Hall resistance was measured under an out-of-plane external field ($H_{ext,z}$) sweeping between −40 and 40 kOe. The values of both $R_{AHE}$ and $H_k$ are estimated from the plot of $R_{xy}$ against $H_{ext,z}$. The CoFeB layer is in-plane magnetized as shown by the $R_{xy}$ against $H_{ext,z}$ in Fig. 3(b). Linear fits were performed to the high-field data to eliminate the ordinary Hall effect. $R_{AHE}$ is obtained from the intercept of the fitted curve. $H_k$ is determined by the magnetic field for saturating $R_{xy}$ after subtracting the contribution of the ordinary Hall effect. The harmonic Hall measurement was carried out by applying a current density of the order of $j_{CoSi}$ ~ 1.6×10$^6$ A/cm$^2$ in CoSi. Figure 3(c) shows the $\varphi$ dependence of $R_\omega$ for a Hall device with $t_{CoSi}$ = 7.2 nm at $H_{ext}$ = 2 kOe. Since $M$ lies in the $xy$ plane ($\theta$ = 90°), the $R_\omega$-$\varphi$ curve is well fitted by Eq.(2). The $R_{PHE}$ is obtained to be 65 mΩ. $R_{2\omega}$ as a function of $\varphi$ for same the device is plotted in Fig. 3(d). The $A\cos\varphi$ and $B\cos 2\varphi\cos\varphi$ components of $R_{2\omega}$ are extracted by fitting the experimental data using Eq. (3). The factor $A$ as a function of $H_{ext}$ for the device with $t_{CoSi}$ = 7.2 nm is plotted in Fig. 3(e). The best fit using Eq. (3) is shown by the red line and other colored lines represent decomposition of the signal. The SOT signal dominates in the low-field regime whereas the $H_{ext}$ dependence of $R_{2\omega}$ at higher fields is governed by the ONE. We note that previous report [28] has found appreciable ONE in CoSi single crystal at lower temperatures. The $H_{DL}$ is estimated to be 1.00 Oe. Figure 3(f) shows the dependence of the factor $B$ against the inverse of $H_{ext}$. Combining with the $R_{PHE}$ obtained above and the $H_{Oe}$ evaluated by the Ampère's law, the $H_{FL}$ opposing to the Oersted field is estimated to be 2.25 Oe from the slope of the linear curve fit in Fig. 3(f). The $\xi_{DL}$ and $\xi_{FL}$ are given by [35],

$$\xi_{DL\,(FL)} \equiv \frac{2e}{\hbar} \frac{H_{DL\,(FL)} M_s t_{CoFeB}}{j_{CoSi}} \qquad (4)$$

where $e$ is the elementary charge, $\hbar$ is the reduced Plank constant, and $M_s = 980$ emu/cm$^3$ is the saturation magnetization of CoFeB. The CoSi thickness dependences of both $\xi_{DL}$ and $\xi_{FL}$, are shown in Fig. 3(g). The result is comparable to the spin Hall efficiency estimated by SMR measurement. The increase in the $\xi_{DL}$ and $\xi_{FL}$ for the 4.5-nm-thick CoSi sample could be due to the contribution from the surface roughness [36]. Based on the definition of $\sigma_{DL(FL)} \equiv \xi_{DL(FL)}/\rho_{CoSi}$, the equivalent spin Hall conductivities of damping like and field like SOT, i.e. $\sigma_{DL}$ and $\sigma_{FL}$, are plotted in Fig. 3(h). We further fitted the thickness dependence of $\sigma_{DL(FL)}$ using the equation [37],

$$\sigma_{DL(FL)}(t) \equiv \sigma_{DL(FL)} \left[1 - \mathrm{sech}(t_{CoSi}/\lambda)\right] \qquad (5)$$

and finally, with $\lambda = 6$ nm, we obtained $\sigma_{DL} = 45$ $(\hbar/e)\Omega^{-1}\mathrm{cm}^{-1}$ and $\sigma_{FL} = 95$ $(\hbar/e)\Omega^{-1}\mathrm{cm}^{-1}$.

The temperature dependence of the magneto-transport for single-layer CoSi was systematically studied. The inset of Fig. 4(a) shows the temperature $T$ dependence of the normalized resistivity. We found the temperature coefficient of the resistivity for CoSi is negative and $\rho_{CoSi}$ increases by ~10% as $T$ is swept from 300 K to 10 K. We then extract the carrier concentration $n$ and the electron mobility $\mu$ of CoSi from the slope of the ordinary Hall effect and the longitudinal conductivity. Results are plotted in Fig. 4(a). The temperature dependences of the two quantities are relatively weak. We next investigate the $T$ dependence of the charge-to-spin conversion for a typical CoSi/CoFeB bilayer stack with $t_{CoSi} = 8.5$ nm using the harmonic Hall technique. $H_{ext}$ dependence of $A$ measured at various temperatures are plotted in Fig. 4(b). We found both the damping-like SOT contribution (exponential decay of $A$ at low field) and ONE contribution (linear slope of $A$ at high field) of $R_{2\omega}$ reduce with decreasing $T$. At 10K, $R_{2\omega}$ is nearly independent of $H_{ext}$. Figure 4(c) shows $B$ as a function of $1/H_{ext}$ at various temperatures. The slope of $B$ changes sign which indicates a strong temperature dependence of $\xi_{FL}$. At 10K, the current-induced Oersted field is sufficient to explain the observed signal, implying that $\xi_{FL}$ is nearly zero. Figure 4(d) plots the $T$ dependence of $\xi_{DL}$ and $\xi_{FL}$. Both $\xi_{DL}$ and $\xi_{FL}$ decrease by almost one order of magnitude with decreasing temperature and nearly vanish at 10 K. The $T$ dependences of $\sigma_{DL}$ and $\sigma_{FL}$ shown in Fig. 4(e) also exhibit similar trend. While the $\xi_{FL}$ of heavy metal/CoFeB/MgO-based heterostructures was known to show such a strong temperature dependence, the $T$ dependence of the damping-like counterpart in those structures is relatively weak [38,39].

The scaling relationship of the damping-like spin Hall conductivity $\sigma_{DL}$ against the transport lifetime $\tau$ is commonly used to separate the intrinsic (independent of $\tau$) and the extrinsic skew scattering (proportional to $\tau$) contributions [40]. For metals, it is convenient to assume $\sigma_{xx} \sim \tau$. However, for semimetallic CoSi, since $n$ varies with temperature (Fig. 4(a)), we plot $\mu$ instead of $\tau$ in the $x$-axis in Fig. 4(f). For the $y$-axis, we assume $\sigma_{DL}$ may scale with $n$ due to the thermal excitation [41]. Interestingly, the scaling analyses (whether we divide $\sigma_{DL}$ by $n$ or not) show a strong temperature dependence of $\sigma_{DL}$ in CoSi, which neither follows usual intrinsic nor extrinsic skew scaling [40,42,43], as shown in Fig. 4(f). Thermal-excitation-related extra extrinsic scatterings and coupling, such as the local moments [44] of Co ions, and phonons [45], as well as the shift of Fermi level with its special electronic structures [46], may contribute to the charge-spin conversion in CoSi at elevated temperatures.

In order to shed light on the mechanism of the spin-current generation in CoSi thin films, first-principles calculations based on the density-functional theory (DFT) [47–49] were carried out to evaluate the intrinsic SHC in the *B*20-type CoSi. The relativistic band structures along high-symmetry ***k*** paths are presented in Fig. 5(a). Band crossings with high-fold degeneracy are confirmed at Γ (near the Fermi energy) and R (at 0.2 eV below the Fermi energy) in the Brillouin zone (BZ) as reported previously [19–21]. These states were identified as the spin-1 chiral fermions and the double Weyl fermions, respectively [18]. The degeneracy of these states, as shown in Fig. 5(a) with inset, is reduced by the spin-orbit coupling (SOC) and the split bands might be an origin of the enhanced SHC in CoSi. Figure 5(b) (black line) presents the contributions of the spin Berry curvature along the corresponding high-symmetry ***k*** paths at the Fermi energy. A remarkable enhancement of the spin Berry curvature arises near the Γ point depending on the particular directions, namely Γ – X, Γ – X', and Γ – M directions. The largest contribution is found in the vicinity of Γ in Γ – X line, where X is at (0.5, 0, 0). In contrast to Γ – X, both positive and negative contributions can be seen at Γ – X' line, where X' is at (0, 0.5, 0). Smaller value of $\Omega_{xy}^z$ is obtained in Γ – M line. The calculations obtain a SHC of ~52 ($\hbar/e$) $\Omega^{-1}\text{cm}^{-1}$ on the Fermi level. Figure 5(c) shows the energy dependence of the SHC within the rigid band model, indicating the value and sign of the SHC depend on the Fermi energy. The maximum amplitudes of the SHC are found to be ~147 and ~119 ($\hbar/e$) $\Omega^{-1}\text{cm}^{-1}$ when the Fermi energy shifts down to −0.16 and up to 0.24 eV, respectively. For the former (at −0.16 eV), it is confirmed that the spin Berry curvature is enhanced near the bands crossing of double Weyl point at R in BZ. Therefore, controlling the Fermi energy by doping with other elements could be an effective way to achieve a larger SHC in this material.

Interestingly, SHC vanishes at the spin-1 band crossing and changes its sign above and below the topological node, whereas rather a peak-like structure of SHC was observed around the double Weyl point (Fig. 5(c)). The distinct odd-function-like energy dependence of SHC for the spin-1 chiral fermion clearly distinguishes itself from others (e.g. massive Dirac fermion [50], inversion asymmetric Weyl fermion [12], magnetic Weyl fermion [51], and Dirac fermion from the topological surface states [46]) where the Fermi energy dependence of the (anomalous and spin) Hall conductivities are rather even-like.

To provide more insights into this unique character, we show the contributions of $\Omega_{xy}^z$ on the slices of BZ ($k_x k_y$ plane at $k_z=0$) at varying energies of 0.24, 0.12, 0.06, 0.03, 0, and −0.03 eV in Figs. 6(a)−6(f), respectively. At the spin-1 chiral band crossing point (0.03 eV; Fig. 6(d)), the positive contribution of $\Omega_{xy}^z$ is found around Γ point with two-fold symmetry. When the energy shifts upward to 0.06 eV (Fig. 6(c)), the negative $\Omega_{xy}^z$ appears along Γ – X' line. With further increasing the energy to 0.12(Fig. 6(b)) and to 0.24 eV (Fig. 6(a)), the negative contribution of $\Omega_{xy}^z$ is increased and eventually, the value of SHC reaches to the minimum value, −119 ($\hbar/e$) $\Omega^{-1}$cm$^{-1}$, at 0.24 eV. On the other hand, the $\Omega_{xy}^z$ values increase as lowering energy to the Fermi level (0 eV) and to −0.03 eV (Figs. 6(e) and 6(f)), resulting in the positive SHC below the spin-1 chiral bands.

It is found that the value of the spin Berry curvature can be understood in terms of the *d–p* orbital hybridization between nearest Co and Si atoms through the electric dipole transition from occupied *d* states to unoccupied *p* states. In Fig. 5(a), orbital weights of Co *d* states are dominant below the spin-1 chiral fermions and these orbitals hybridize with a small component of Si *p* orbital. The hybridization of Co *d* and Si *p* orbitals plays an important role in determining the SHC in the CoSi. According to the selection rule for the electric dipole transitions [52], the transitions from occupied *d* state of $|\mathbf{k}lmn\rangle$ with angular quantum number $l = 2$ to unoccupied *p* state of $\langle \mathbf{k}l'm'n'|$ with $l' = 1$ that satisfy magnetic quantum number $m = m' - 1$ ($m = m' + 1$) may give a positive (negative) contribution to the SHC (*n* is a band index). Below the spin-1 chiral fermions, our calculations clarified that the positive value of SHC originates from the transitions satisfying $l = l' + 1$ and $m = m' - 1$, specifically, from $d_{-1}$ state ($l = 2, m = -1$) to $p_0$ state ($l' = 1, m' = 0$) and from $d_{-2}$ state to $p_{-1}$ state along Γ – X' line, and $d_0$ state to $p_1$ state along Γ – X line, respectively. Note, the transitions $d_1 \rightarrow p_0$ and $d_2 \rightarrow p_1$ that give negative contributions are also confirmed around the Fermi level. These transitions cancel a part of the positive contributions, thus, the resulting value of SHC in CoSi is relatively small at the Fermi level. In the higher energy region ($E > 0.03$ eV), the transitions $d_0 \rightarrow p_{-1}$ were found to give the negative SHC. We mention that above and below the double Weyl point, the positive $\Omega_{xy}^z$ is attributed to $d_{-1} \rightarrow p_0$ transitions around R point of $\Gamma - R$ line.

We further emphasize that the small Si component hybridizing with Co $d$ states is an essential ingredient for the SHC in CoSi. Further DFT calculations were carried out for $B$20-CoSi where the SOC of either Co or Si is turned off. Since the SOC in Si is very weak, the calculations with SOC only for Si obtain a tiny value of the SHC, $-1$ ($\hbar/e$) $\Omega^{-1}$cm$^{-1}$ (magenta line in Fig. 5(c)). In contrast, the inclusion of the SOC only for Co raises the SHC up to ~38 ($\hbar/e$) $\Omega^{-1}$cm$^{-1}$ (cyan line in Fig. 5(c)); however, this value is much smaller than the SHC of CoSi, where the SOC of both atoms are turned on. Particularly, the peak of spin Berry curvature along $\Gamma - X$ line is doubled for the full SOC case as shown in Fig. 5(c). Thus, the bonding of Si $p$ states with Co $d$ states plays a significant role in the SHC despite the "small" (~1%) amount of the Si $p$ character in the hybridized band structures.

As a final remark, the vanishment of SHC contribution at the spin-1 chiral crossing near the Fermi level results in the relatively small $\xi_{DL}$ observed in this study. The SHC exhibits an odd-function-like energy dependence for the spin-1 chiral fermion. The sign of SHC changes above and below the topological node and the amplitude of SHC is close to zero at the spin-1 chiral crossing. Since the spin-1 chiral crossing locates very near the Fermi level, the $\xi_{DL}$ proportional to SHC is small. The local derivative in energy of SHC $\partial \sigma_{xy}/\partial E$ is however robust near the topological node and may be exploited for thermal spin current generation via the spin Nernst effect [53,54]. Furthermore, it would be interesting to investigate the isostructural $B$20 PtAl [55] for which dramatic enhancement of SOC is expected to boost up $\partial \sigma_{xy}/\partial E$ at the crossing and peaks of $\sigma_{xy}$ slightly away from the crossing.

In summary, the spin-1 chiral CoSi semimetal thin films and heterostructures were fabricated for quantifying the SOTs. The CoSi films grown on sapphire *c*-plane substrates showed a *B*20 crystal structure in polycrystalline phase with flat surface morphology. The spin-current generation in the CoSi films via SHE was investigated by SMR and harmonic Hall measurements. The spin Hall efficiency of the CoSi films was evaluated to be ~3% at room temperature and reduced with decreasing temperature. First-principles calculations indicate that the hybridization between *d-p* orbitals results in a large enhancement of spin Berry curvature, which mainly contributes to the SHC in the CoSi. The unique antisymmetric-like energy dependence of SHC highlight the critical role of Fermi level tuning for harnessing all the benefits of exotic spin-1 chiral fermions for spin current generation driven by either an electric field or a thermal gradient.

**Acknowledgements**

K. Tang acknowledges National Institute for Materials Science for the provision of a NIMS Junior Research Assistantship. We thank Kohji Nakamura at Mie University and Guanxiong Qu at University of Tokyo for fruitful discussions. This work was partially supported by the KAKENHI (No. JP20K04569, No. JP20H02190, No. JP20H00299, JP16H06332, and No. JP20K15156) from the Japan Society for the Promotion of Science (JSPS), the Inter-University Cooperative Research Program of the Institute for Materials Research, Tohoku University (No. 20K0058).

K. Tang, Y.L. and K.N. contributed equally to this work.

**APPENDIX: MATERIALS AND METHODS**

**1. Film deposition, characterizations, and device fabrication**

The CoSi thin films were deposited using magnetron sputtering at the substrate temperature of 550°C in a high-vacuum sputter chamber. The base and deposition pressures were $2 \times 10^{-6}$

Pa and 0.7 Pa (Ar gas), respectively. The sputtering power of CoSi was set at 20 W generated by a direct current (dc) source for a 76.2 mm diameter $Co_{50}Si_{50}$ target. The crystal structure was characterized using the out-of-plane XRD measurement with Cu $K_\alpha$ radiation ($\lambda = 0.15418$ nm). The surface structure and morphology were evaluated using RHEED and AFM, respectively. The sample with the structure of sapphire(0001) substrate/CoSi (0~11 nm)/CoFeB (1 nm)/MgO (2 nm)/Ta (1 nm) was post annealed at 350°C for 1 h and was then patterned into Hall bar devices (width: 10 μm, length: 25 μm) by conventional UV lithography and Ar ion milling. Ta(5)/Au(150 nm) layers were then deposited on the Hall bars as electrodes using lift-off process.

**2. SMR and harmonic Hall measurements**

Magneto-transport properties were characterized in a physical properties measurement system (PPMS) at room temperature. For the SMR measurement, a magnetic field of 20 kOe was applied to saturate the magnetization of the FM layer. For the harmonic Hall measurement, a sinusoidal signal of constant amplitude and frequency of 172.1 Hz was applied by a Keithley 6221 current source meter. The first- and second-harmonic Hall voltages were simultaneously measured by two lock-in amplifiers (nf LI5660).

**3. The first-principles calculations**

The first-principles calculations were performed based on the generalized gradient approximation [56] using the full-potential linearized augmented plane-wave (FLAPW) method [47–49]. The cubic *B*20-type crystal structure of stoichiometric CoSi was modelled by using the experimental lattice constant, in which all the atomic positions were fully relaxed by the force calculations [57]. Muffin-tin (MT) radii of 2.40 and 1.80 $a_B$ were employed for Co and Si, respectively. The LAPW basis for the wave function in the interstitial region has a cutoff of $|\mathbf{k} + \mathbf{G}| \leq 4.5\ a_B^{-1}$, and the angular momentum expansion inside the MT sphere is truncated at $\ell = 8$ for the Co and at 6 for the O. The SOC was incorporated by the second variational method [58]. The intrinsic SHC was evaluated by means of the Kubo formula in the static limit ($\omega = 0$) [59,60],

$$\sigma_{xy} = \frac{e}{\hbar}\sum_{\boldsymbol{k}} \sum_{n} f_{\boldsymbol{k}n}\Omega^z(\boldsymbol{k}),$$

where the spin Berry curvature $\Omega^z(\boldsymbol{k})$ is given by

$$\Omega^z(\boldsymbol{k}) = \sum_{n'} \frac{2\mathrm{Im}\langle \boldsymbol{k}n|\hat{\jmath}_x|\boldsymbol{k}n'\rangle\langle \boldsymbol{k}n'|\hat{v}_y|\boldsymbol{k}n\rangle}{(\varepsilon_{\boldsymbol{k}n}-\varepsilon_{\boldsymbol{k}n'})^2}.$$

Here, $f_{\boldsymbol{k}n}$ is the Fermi distribution function for the *n*-th band at $\boldsymbol{k}$, $\hat{\jmath}_x$ and $\hat{v}_y$ are the spin and charge current operators, and $\varepsilon_{\boldsymbol{k}n}$ and $\varepsilon_{\boldsymbol{k}n'}$ are the calculated eigenvalues of the occupied and unoccupied states, $|\boldsymbol{k}n\rangle$ and $|\boldsymbol{k}n'\rangle$, respectively. With a $55 \times 55 \times 55$ special $\boldsymbol{k}$-point mesh in the BZ, the SHC was found to sufficiently suppress numerical fluctuations.

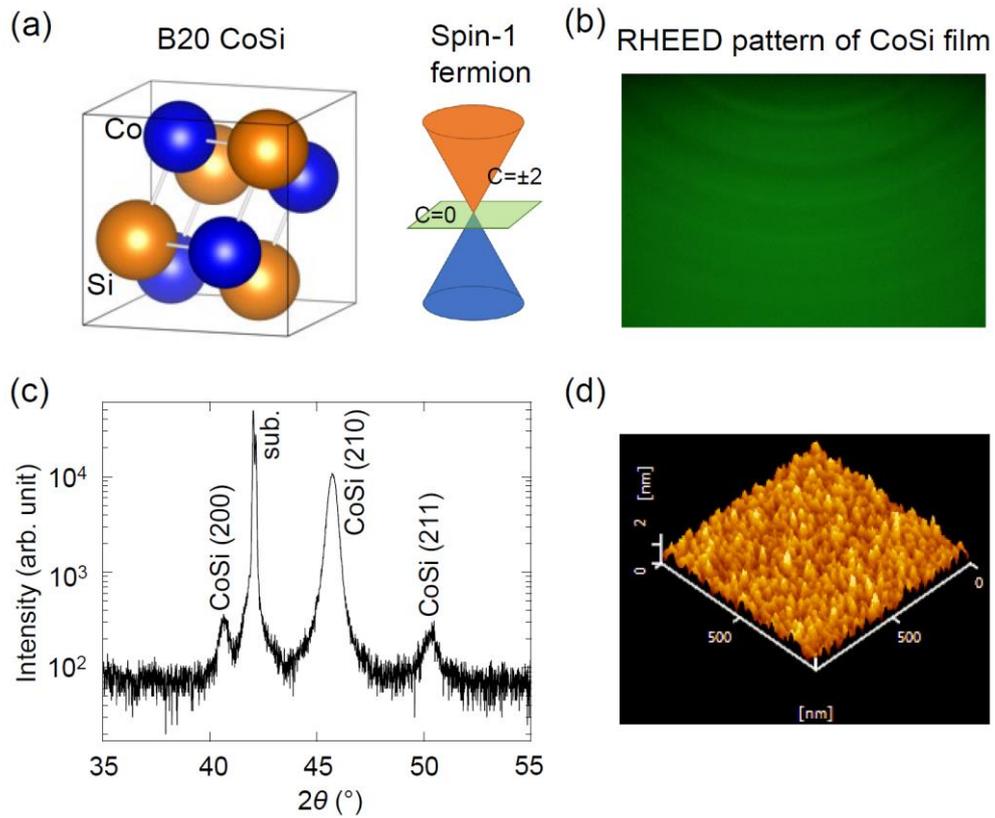

FIG. 1. Structural properties of CoSi films and SOT devices. (a) Unit cell of the cubic *B*20 crystal structure of CoSi and schematic of the band structures of a spin-1 fermion. (b) RHEED pattern of CoSi (47 nm) film surface deposited on a sapphire *c*-plane substrate. (c) The out-of-plane XRD spectrum of substrate/CoSi (47)/Mg$_2$AlO$_x$ (2 nm). (d) The surface morphology of multilayer stacks with the structure of CoSi/CoFeB/MgO for SOT devices.
19

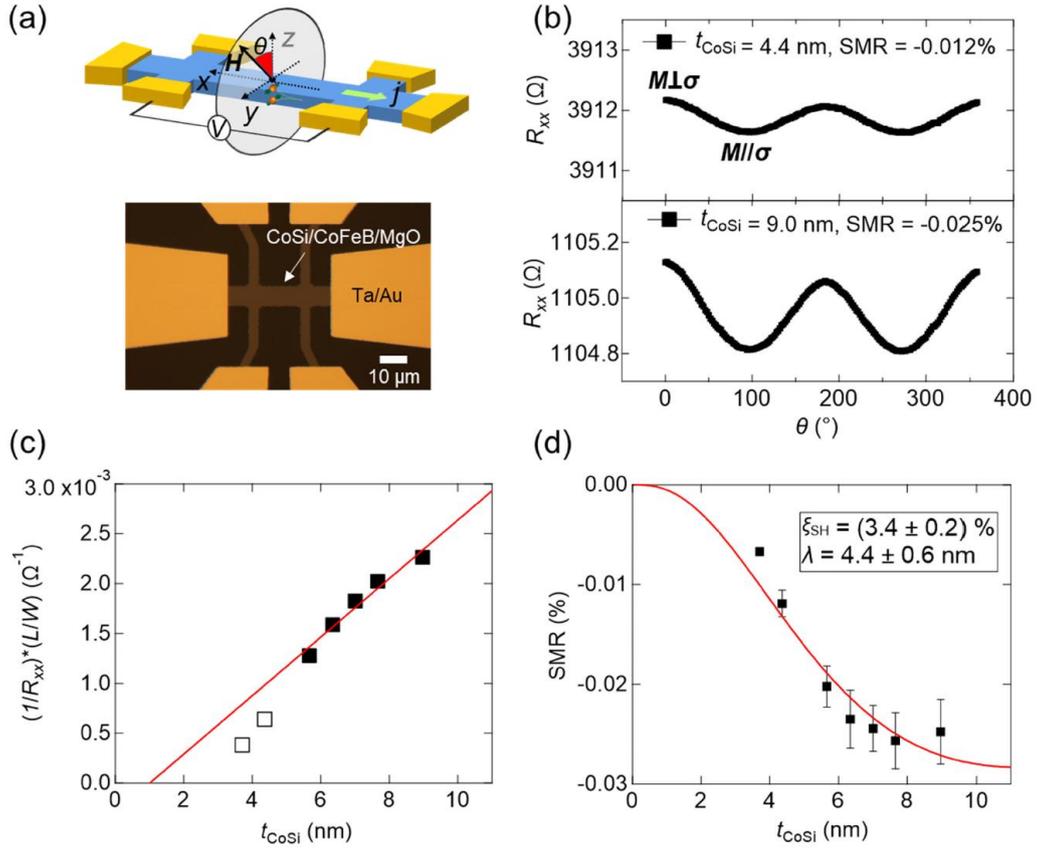

FIG. 2. The measurement of SMR in CoSi. (a) Illustration of a Hall bar device with the measurement setup and coordinate system, and an optical microscope image of a Hall bar device. (b) Plots of the dependence of resistance on the angle between $M$ and $\sigma$ for two Hall bar devices with $t_{CoSi}$ = 4.4 and 9.0 nm. (c) Inverse of the sheet resistance as a function of CoSi thickness. The red solid line represents the linear fitting curve in a proper range shown by the black solid squares. (d) Extracted SMR ratio as a function of CoSi thickness. Red solid line is the fitting curve using Eq. (1).

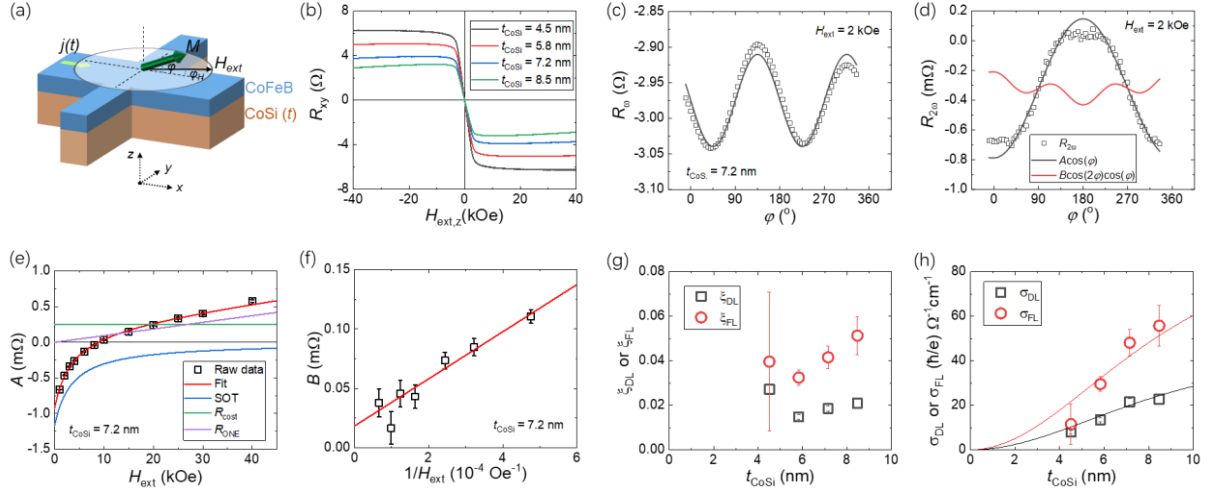

FIG. 3. Harmonic Hall measurement for the SOTs generated from CoSi films. (a) Schematic of the setup for the harmonic Hall measurement. (b) Hall resistance measured under an out-of-plane magnetic field for Hall bar devices with $t_{CoSi}$ = 4.5, 5.8, 7.2, and 8.5 nm. (c) First harmonic Hall resistance $R_\omega$ as a function of $\varphi$. Solid line is the fitted curve by Eq. (2). (d) Second harmonic Hall resistance $R_{2\omega}$ as a function of $\varphi$. Solid curves are the decomposed components of $R_{2\omega}$ based on Eq. (3). $R_\omega$ and the $R_{2\omega}$ shown in (c) and (d) were measured at $H_{ext}$ = 2 kOe for the device with $t_{CoSi}$ = 7.2 nm. (e) Parameters $A$ and (f) $B$ as a function of $H_{ext}$ and the inverse of $H_{ext}$, respectively. (g) CoSi thickness dependence of $\xi_{DL}$ and $\xi_{FL}$. (h) $\sigma_{SH}$ and $\sigma_{FL}$ as a function of CoSi thickness. The solid lines are the fitting curves by Eq. (5).

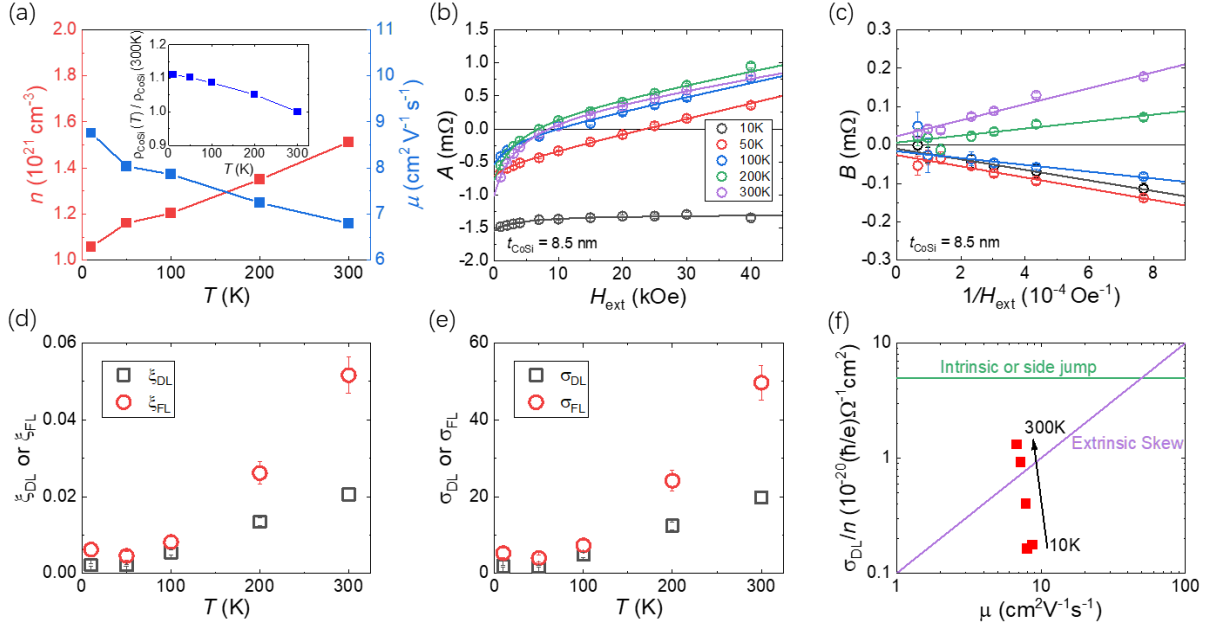

FIG. 4. Temperature dependence of transport properties and SOTs for CoSi films and CoSi/CoFeB heterostructures. (a) The dependence of carrier concentration, mobility, and resistivity on temperature. (b) Parameters $A$ and (c) $B$ as a function of $H_{ext}$ and the inverse of $H_{ext}$, respectively, at different temperatures. (d) Temperature dependences of $\xi_{DL}$ ($\xi_{FL}$) and (e) $\sigma_{SH}$ ($\sigma_{FL}$). (f) $\sigma_{DL}/n$ as a function of the carrier mobility for the CoSi/CoFeB heterostructures.

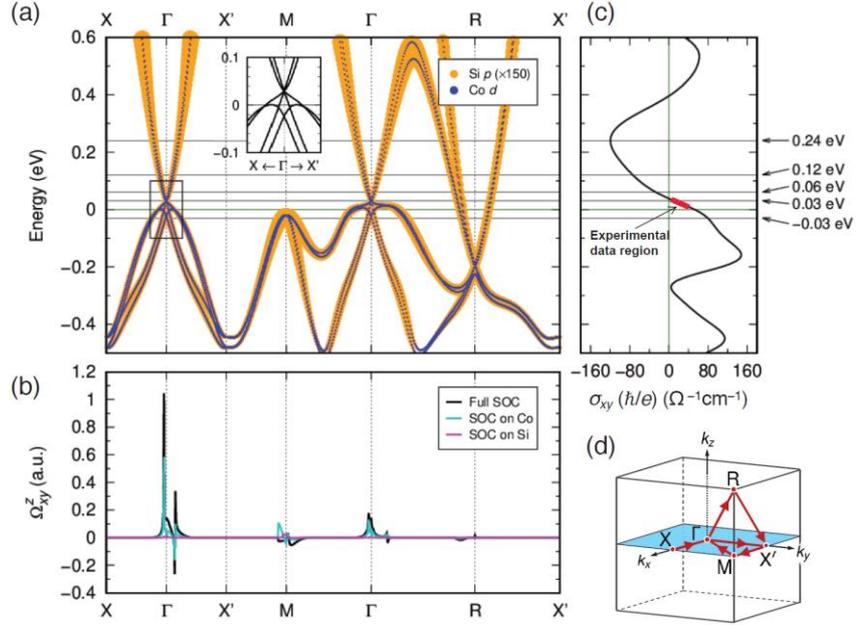

FIG. 5. Band structure and spin Hall conductivity of CoSi. (a) Relativistic band structure of $B$20-CoSi along high-symmetry $k$ points in the first BZ. Weights of the Co $d$ (blue) and Si $p$ (yellow) orbitals are represented by size of the symbol. The Fermi level is at $E = 0$ eV. Horizontal lines indicate the energy positions relative to the Fermi energy. Inset focuses on the crossing bands of spin-1 chiral fermion around Γ. (b) Spin Berry curvature ($\Omega_{xy}^z$) contributions along the high-symmetry $k$-point path on the Fermi level. The results from the full SOC (black) and the SOC on only Co (cyan) or Si (magenta) are shown. (c) Energy dependence of calculated spin Hall conductivity, $\sigma_{xy}$, in the rigid band model. The region of experimental results is also indicated. (d) Illustration of the first BZ showing the $k$ paths. Note that X is at (0.5, 0, 0) and X' is at (0, 0.5, 0), respectively.

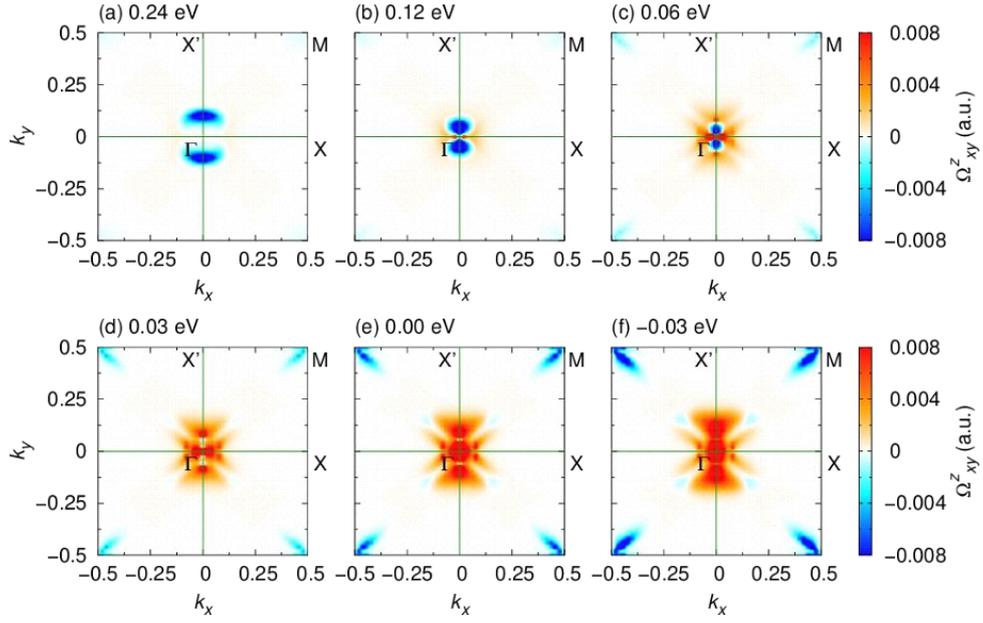

FIG. 6. Spin Berry curvature contributions projected onto $k_x k_y$ plane at $k_z = 0$ in the BZ. (a) Energy slice at 0.24 eV, (b) 0.12 eV, (c) 0.06 eV, (d) 0.03 eV, (e) 0.00 (Fermi level), and (f) −0.03 eV, respectively (see also Fig. 5(a) and 5(b)). The color bar indicates positive (red) and negative (blue) contributions of $\Omega_{xy}^z$. The $k_x k_y$ plane corresponds to filled area (blue) in Fig. 5(d).